\documentclass [11pt]{article}

\usepackage {graphicx}
\usepackage {longtable}
\begin{document}

\begin{center}
\bf{\LARGE Scalar-Isovector $\delta$-Meson in the Relativistic
Mean-Field Theory and the Structure of Neutron Stars with a Quark
Core}\footnote{Talk given at the International Conference
RUSGRAV-13, June 23-28, 2008, PFUR, Moscow }

\vskip10pt{\bf{Grigor Alaverdyan }
\vskip5pt{\textit {\small
Yerevan State University, A.Manoogyan str. 1, 0025 Yerevan,
Armenia
\\ E-mail: galaverdyan@ysu.am}}}
\end{center}

\begin{quote}
{\small \textbf{Abstract.} In the framework of the relativistic
mean-field theory, we have considered the equation of state of
superdense nuclear matter, taking into account an effective
scalar-isovector $\delta$-meson field. The effect of the
$\delta$-meson field on the characteristics of a Maxwell-type
quark phase transition has been studied. The quark phase is
described with the aid of the improved version of the MIT
(Massachusetts Institute of Technology) bag model, in which
interactions between the $u,~ d,~ s$ quarks inside the bag are
taken into account in the one-gluon exchange approximation. For
different values of the bag parameter $B$, series of neutron star
models with a quark core have been built. Stability problems for
neutron stars with an infinitesimal quark core are discussed. An
estimate is obtained for the amount of energy released in a
catastrophic transformation of a critical neutron star to a star
with a finite-size quark core.

\textbf{Key words:} Equation of state, mean-field, neutron stars, deconfinement phase transition\\
\textbf{PACS numbers:} 97.60.Jd, 26.60.+c, 12.39.Ba }
\end{quote}

\section{Introduction}

The properties of such compact objects as neutron stars (NS)
depend functionally on the equation of state (EOS) of matter in a
sufficiently widerange of densities. A knowledge of the structure
characteristics and the constituent  composition of matter at
extremely large densities is a necessary condition for revealing
the physical nature of the NS internal structure and integral
parameters. One of the effectively used theories, sufficiently
adequately describing the properties of nuclear matter as a system
of strongly interacting baryons and mesons, is the relativistic
mean field (RMF) theory [1, 2]. This theory has made possible to
obtain results which satisfactorily describe the structure of
finite nuclei [3], the EOS of nuclear matter [4], and the features
of heavy ion scattering [5]. Inclusion of a scalar-isovector
$\delta$ meson into the scheme and a study of its role for
asymmetric nuclear matter in the low density range has been
conducted in [6, 7, 8]. An objective of this paper is to study the
EOS of superdense nuclear matter in the framework of the RMF
theory and an investigation of changes in the parameters of the
first-order phase transition due to inclusion of $\delta$ meson
exchange. Using the EOS obtained, we calculate the integral and
structure characteristics of NS with a quark core.

\section{Properties of Neutron Star Matter}

\subsection{\textit{ Nuclear Matter }}

The nonlinear Lagrangian density of an interacting multiparticle
system consisting of nucleons and an isoscalar-scalar $\sigma$
meson, an isovector-scalar $\delta$ meson, an isoscalar-vector
$\omega$ meson and an isovector-vector $\rho$ meson as well as
free electrons has the following form in quantum hadrodynamics
(QHD)\footnote{We will use the natural system of units
$\hbar=c=1$}:
\begin{eqnarray}{ \cal L}=\bar {\psi} _{N}\left[ \gamma ^{^{\mu }}\left(
i\partial _{_{\mu }}-g_{\omega }\omega _{_{\mu }}(x)-\frac{1}{2}g_{\rho }%
\overrightarrow{\tau }_{_{N}}\overrightarrow{\rho }_{_{\mu
}}(x)\right) -\left( m_{_{N}}-g_{_{\sigma }}\sigma (x)-g_{_{\delta
}}\overrightarrow{\tau }_{_{N}}\overrightarrow{\delta }(x)\right)
\right] \psi _{N} \nonumber \\
+\frac{1}{2}\left( \partial _{_{\mu }}\sigma (x)\partial ^{^{\mu
}}\sigma (x)-m_{_{\sigma }}\sigma (x)^{2}\right)
-\frac{b}{3}m_{_{N}}\left( g_{_{\sigma }}\sigma (x)\right)
^{3}-\frac{c}{4}\left( g_{_{\sigma }}\sigma
(x)\right) ^{4}  \nonumber \\
+\frac{1}{2}m_{_{\omega }}^{2}\omega ^{^{\mu }}(x)\omega _{_{\mu }}(x)+%
\frac{1}{2}m_{_{\rho }}^{2}\overrightarrow{\rho }^{^{\mu }}(x)%
\overrightarrow{\rho }_{_{\mu }}(x)-\frac{1}{4}\Omega _{_{\mu \nu
}}(x)\Omega ^{^{\mu \nu }}(x)-\frac{1}{4}R_{_{\mu \nu
}}(x)R^{^{\mu \nu
}}\left( x\right) \nonumber \\
+\frac{1}{2}\left( \partial _{_{\mu }}\overrightarrow{\delta
}(x)\partial
^{^{\mu }}\overrightarrow{\delta }(x)-m_{_{\delta }}^{2}\overrightarrow{%
\delta }(x)^{2}\right) + \bar {\psi} _{e}\left( i\gamma ^{^{\mu
}}\partial _{_{\mu }}-m_{_{N}}\right) \psi _{e}
\end{eqnarray}
Here, $g_{\sigma}, \;g_{\omega}, \;g_{\delta}$ and $g_{\rho}$
denote the coupling constants of the nucleon with the
corresponding meson. In the RMF theory, the meson fields
$\sigma\left({x}\right)$, $\omega_{\mu}\left({x}\right)$,
$\vec{\delta}\left({x}\right)$ and $\vec{\rho}_{\mu}
\left({x}\right)$ are replaced by the (effective) fields $\bar
{\sigma} ,\;\,\bar{\omega} _{\mu} ,\;\,\bar {\vec {\delta
}},\;\bar {\vec {\rho} }_{\mu}$. Re-denoting the meson fields and
coupling constants according to

\begin{equation}
\label{eq1} g_{\sigma}  \bar {\sigma}  \equiv \sigma \,,\quad
g_{\omega}  \bar {\omega }_{0} \equiv \omega \,,\quad g_{\delta}
\bar {\delta} ^{\left( {3} \right)} \equiv \delta \,,\quad \quad
g_{\rho}  \bar {\rho} _{0}^{\left( {3} \right)} \equiv \rho \,,
\end{equation}

\begin{equation}
\label{eq2} \left( {{{g_{\sigma} }  \mathord{\left/ {\vphantom
{{g_{\sigma} } {m_{\sigma} } }} \right. \kern-\nulldelimiterspace}
{m_{\sigma} } }} \right)^{2} \equiv a_{\sigma}  \,,\quad \left(
{{{g_{\omega} } \mathord{\left/ {\vphantom {{g_{\omega} }
{m_{\omega} } }} \right. \kern-\nulldelimiterspace} {m_{\omega} }
}} \right)^{2} \equiv a_{\omega} \,,\quad \left( {{{g_{\delta} }
\mathord{\left/ {\vphantom {{g_{\delta} } {m_{\delta} } }} \right.
\kern-\nulldelimiterspace} {m_{\delta} } }} \right)^{2} \equiv
a_{\delta}  \,,\quad \left( {{{g_{\rho} } \mathord{\left/
{\vphantom {{g_{\rho} }  {m_{\rho} } }} \right.
\kern-\nulldelimiterspace} {m_{\rho} } }} \right)^{2} \equiv
a_{\rho} \,\quad
\end{equation}
and introducing the asymmetry parameter
\begin{equation}\label{eq3} \alpha = {{\left( {n_{n} - n_{p}}  \right)}
\mathord{\left/ {\vphantom {{\left( {n_{n} - n_{p}}  \right)}
{n}}} \right. \kern-\nulldelimiterspace} {n}}\;,
\end{equation} one can present the field equations in the form
\begin{equation}
\label{eq4} \sigma = a_{\sigma}  \left( {n_{s\,p} \left(
{n,\alpha}  \right) + n_{s\,n} \left( {n,\alpha}  \right)\, -
bm_{N} \sigma ^{2} - c\sigma ^{3}} \right),
\end{equation}
\begin{equation}
\label{eq5} \omega = a_{\omega}  n\,,
\end{equation}
\begin{equation}
\label{eq6} \delta = a_{\delta}  \left( {n_{s\,p} \left(
{n,\alpha}  \right) - n_{s\,n} \left( {n,\alpha}  \right)\,}
\right),
\end{equation}
\begin{equation}
\label{eq7} \rho = - \frac{{1}}{{2}}a_{\rho}  \,n\,\alpha \;,
\end{equation}
\noindent where \begin{equation} \label{eq8} n_{s\,p} \left(
{n,\alpha} \right) = \frac{{1}}{{\pi
^{2}}}\int\limits_{0}^{k_{F\,} \left( {n} \right)\left( {1 -
\alpha} \right)^{1/3}} {\frac{{m_{p}^{\ast}  \left( {\sigma
,\delta} \right)}}{{\sqrt {k^{2} + m_{p}^{\ast}  \left( {\sigma
,\delta} \right)\,^{2}}} }} \;k^{2}dk\; \quad ,
\end{equation}

\begin{equation}
\label{eq9} n_{s\,n} \left( {n,\alpha}  \right) = \frac{{1}}{{\pi
^{2}}}\int\limits_{0}^{k_{F\,} \left( {n} \right)\left( {1 +
\alpha} \right)^{1/3}} {\frac{{m_{n}^{\ast}  \left( {\sigma
,\delta} \right)}}{{\sqrt {k^{2} + m_{n}^{\ast}  \left( {\sigma
,\delta} \right)\,^{2}}} }} \;k^{2}dk\; \quad ,
\end{equation}

\begin{equation}
\label{eq10} k_{F} \left( {n} \right) = \left( {\frac{{3\pi
^{2}n}}{{2}}} \right)^{1/3}\; \quad .
\end{equation}
The effective masses of the proton and neutron are determined by
the expressions
\begin{equation}
\label{eq11} m_{p}^{\ast}  \left( {\sigma ,\delta}  \right) =
m_{N} - \;\sigma - \;\delta \;,\quad \quad m_{n}^{\ast}  \left(
{\sigma ,\delta}  \right) = m_{N} - \sigma + \;\delta \,.
\end{equation}
If the constants $a_{\omega}$ and $a_{\rho}$ are known, equations
(6) and (8) determine the functions $\omega \left( {n} \right)$
and $\rho \left( {n,\alpha} \right)$. Moreover, a knowledge of the
other constants $a_{\sigma}$, $a_{\delta}$, $b$, and $c$ makes it
possible to solve the set of equations (5), (7), (9), (10) in a
self-consistent way and to determine the remaining two meson field
functions $\sigma \left( {n,\alpha} \right)$ and $\delta \left(
{n,\alpha} \right)$. The standard QHD procedure makes it possible
to obtain expressions for the energy density $\varepsilon\left(
{n,\alpha} \right)$ and pressure $P\left( {n,\alpha} \right)$ [9].

\begin{table}
\centering
\caption{Constants of the theory without ($\sigma
\omega \rho $) and with ($\sigma \omega \rho \delta$)the $\delta$
meson field} \label{1}

\begin{tabular}{|c|c|c|c|c|c|c|}
\hline & $a_{\sigma }$,\ fm$^{2}$ & $a_{\omega }$,\ fm$^{2}$ &
$a_{\delta }$,\ fm$^{2}$ & $a_{\rho }$,\ fm$^{2}$ & $b$,\
fm$^{-1}$ & $c$ \\ \hline $\sigma \omega \rho $ & $9.154$ &
$4.828$ & $0$ & $4.794$ & $1.654\cdot 10^{-2}$ & $1.319\cdot
10^{-2}$ \\ \hline
$\sigma \omega \rho \delta $ & $9.154$ & $4.828$ & $2.5$ & $13.621$ & $%
1.654\cdot 10^{-2}$ & $1.319\cdot 10^{-2}$ \\ \hline
\end{tabular}%
\end{table}

The constants of the theory are numerically defined in such a way
as to reproduce the empirically known values of such nuclear
characteristics at saturation as the bare nucleon mass $m_N =
938.93$ MeV, the parameter $\gamma =m_{N}^{\ast}/m_{N}= 0.78$, the
saturated nuclear matter baryon number density $n_{0} = 0.153$
fm$^{-3}$ , the binding energy per baryon $f_0 = -16.3$ MeV, the
compressibility modulus $K = 300$ MeV, and the asymmetry energy
$E_{sym}^{\left( {0} \right)} = 32.5$ MeV. To reveal the role of
taking into account the $\delta$ meson, we have used the value
$a_{\delta} =2.5$ fm${^{2}}$ (see [10]). Table 1 shows the values
of constants of the theory obtained without taking into account
the interaction channel due to the isovector scalar $\delta$ meson
($\sigma\omega\rho$) and taking it into account
($\sigma\omega\rho\delta$).

\begin{figure} [ptbh]
\begin{center}
\includegraphics[height=3 in, width=3 in]%
{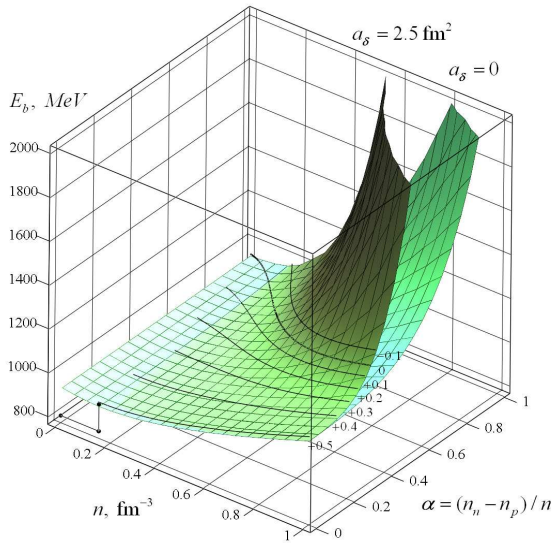}%
\caption{A three-dimensional representation of the energy per
baryon $E_{b}$ as a function of the baryon number density $n$ and
the asymmetry parameter $\alpha$ in the case of a $\beta
$-equilibrium charged ``$ npe$''-plasma. The upper surface
corresponds to the ``$\sigma \omega \rho \delta $'' model, the
lower one to the ``$\sigma \omega \rho $'' model.}  \label{1}
\end{center}
\end{figure}

Fig. 1 represents a three-dimensional picture of the energy per
baryon $E_{b} \left( {n,\alpha}  \right) = {{\varepsilon _{NM}}
\mathord{\left/ {\vphantom {{\varepsilon _{NM}}  {n}}} \right.
\kern-\nulldelimiterspace} {n}}$ vs. the baryon number density $n$
and the asymmetry parameter $\alpha$ in the case of a
$\beta$-equilibrium charged "$npe$"-plasma. The lines correspond
to different fixed values of charge per baryon. The thick line
corresponds to a $\beta$-equilibrium, electrically neutral
"$npe$"-matter. The lower surface corresponds to the
"$\sigma\omega\rho$" model, the upper one to the
"$\sigma\omega\rho\delta$" model. It is seen that inclusion of the
$\delta$ meson field increases the energy per nucleon value, and
this change is enhanced as the nuclear matter asymmetry parameter
increases. The results of our analysis show that the
scalar-isovector $\delta$-field leads to an increased stiffness of
the nuclear matter EOS due to splitting the effective masses of
the proton and the neutron thus increasing the asymmetry energy.
The EOS obtained by us in the normal nuclear density range is
matched to the well known Baym-Bethe-Pethick (BBP) EOS [11].

\subsection{\textit{Strange Quark Matter}}

To describe the quark phase, we have used the improved version of
the MIT (Massachusetts Institute of Technology) bag model [12], in
which interactions between the $u, d, s$ quarks inside the bag are
taken into account in the one-gluon exchange approximation [13].
The quark phase consists of the three quark flavors $u, d, s$ and
electrons in equilibrium with respect to weak interactions. For
the quark masses we have used the values $m_u =5$ MeV, $m_d =7$
MeV and $m_s = 150$ MeV, while for the strong interaction constant
we have taken the value $\alpha_s =0.5$. The calculations were
conducted for different values of the bag parameter $B$ in the
range of 60 to 120 MeV/fm$^3$.

\begin{table}[h]
\centering \caption{Parameters of the Maxwell phase transition for
different values of the "bag" constant $B$ without taking into
account the $\delta$ meson field ($\sigma \omega \rho $)and with
it ($\sigma \omega \rho \delta$).} \label{1}
\begin{tabular}
[c]{|c|c|c|c|c|c|c|c|c|}\hline $B$ &  & $\mu_{b}$ & $n_{N}$ &
$n_{Q}$ & $P_{0}$ & $\varepsilon_{N}$ &
$\varepsilon_{Q}$ & $\lambda$\\
$MeV/fm^{3}$ &  & $MeV$ & $fm^{-3}$ & $fm^{-3}$ & $MeV/fm^{3}$ &
$MeV/fm^{3}$ & $MeV/fm^{3}$ & \\\hline 60 & $\sigma\omega\rho$ &
965.4 & 0.1220 & 0.2826 & 1.965 & 115.8 & 270.9 & 2.299\\\hline 60
& $\sigma\omega\rho\delta$ & 965.9 & 0.1207 & 0.2831 & 2.11 &
114.5 & 271.4 & 2.327\\\hline 69.3 & $\sigma\omega\rho$ & 1037 &
0.246 & 0.3557 & 15.57 & 239.6 & 353.4 & 1.385\\\hline 69.3 &
$\sigma\omega\rho\delta$ & 1032 & 0.2241 & 0.3504 & 13.84 & 217.5
& 347.9 & 1.504\\\hline 80 & $\sigma\omega\rho$ & 1142 & 0.3792 &
0.4819 & 48.54 & 384.5 & 501.8 & 1.159\\\hline 80 &
$\sigma\omega\rho\delta$ & 1119 & 0.3276 & 0.4525 & 37.95 & 328.8
& 468.6 & 1.278\\\hline 100 & $\sigma\omega\rho$ & 1298 & 0.5506 &
0.7175 & 121.3 & 593.4 & 810 & 1.133\\\hline 100 &
$\sigma\omega\rho\delta$ & 1257 & 0.4746 & 0.6497 & 93.30 & 503.3
& 723.5 & 1.213\\\hline 120 & $\sigma\omega\rho$ & 1396 & 0.6512 &
0.8975 & 180.2 & 728.9 & 1073 & 1.18\\\hline 120 &
$\sigma\omega\rho\delta$ & 1354 & 0.5729 & 0.8165 & 143.9 & 631.7
& 961.4 & 1.24\\\hline
\end{tabular}
\end{table}

\subsection{\textit{Phase Transition to Quark Matter under Constant
Pressure}}

We assume in this paper that the nuclear to quark matter
conversion is the ordinary first-order phase transition described
by the Maxwell construction. A separate paper will be devoted to a
study of alterations in the characteristics of a phase transition
with mixed phase formation [14] due to inclusion of the
$\delta$-meson field as well as the effect of these alterations on
the integral and structure parameters of hybrid stars. In the case
of an ordinary first-order phase transition, it is supposed that
both the nuclear and quark matter, taken separately, is
electrically neutral, and at a certain value of the pressure,
$P_0$, corresponding to co-existence of the two phases, the
baryonic chemical potentials of the two phases coincide. Table 2
represents the parameter values of such a phase transition for
five different values of the "bag" parameter $B$ both with and
without the $\delta$ field. In this table, $\mu_b$ is the baryonic
chemical potential at the phase transition point, $n_N$ and $n_Q$
are the baryon number densities of the nuclear and quark matter,
respectively, at the transition point, $\varepsilon_N$ and
$\varepsilon_Q$ are the energy densities, $P_0$ is the phase
transition pressure, and $\lambda= \varepsilon_Q /(\varepsilon_N +
P_0)$ is the density jump parameter. It can be seen that inclusion
of the $\delta$ interaction channel results in decreased values of
the transition pressure $P_0$ and baryon number densities of both
phases at the phase transition point, $n_N$ and $n_Q$. Meanwhile,
the value of the jump parameter $\lambda$ increases.

\begin{figure} [ptbh]
\begin{center}
\includegraphics[height=3in, width=3in]%
{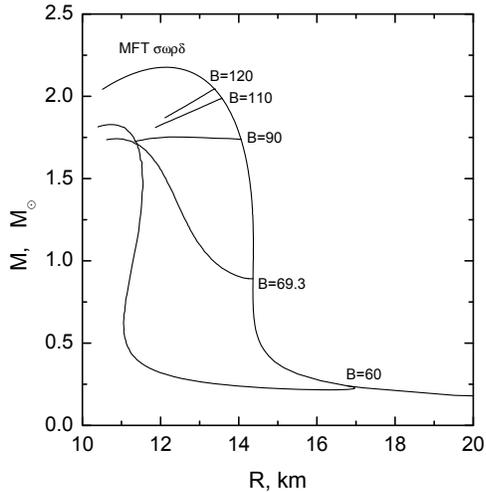}%
\caption{The stellar mass $M$ vs. radius $R$ for different values
of the bag parameter $B$ in the ``$\sigma \omega \rho \delta $''
model. The values of $B$ in MeV/fm$^3$ are shown near the critical
configurations corresponding to quark phase formation.} \label{1}
\end{center}
\end{figure}

\section{Models of Neutron Stars With a Quark Core} Using the EOS
of nuclear matter obtained in the previous section we have
integrated the Tolman-Oppenheimer-Volkoff (TOV) set of equations
[15] and obtained both the structure functions $\varepsilon(r)$,
$P(r)$ and $m(r)$ and the stellar integral parameters: the
gravitational mass $M$ and the radius $R$ for different values of
the central pressure $P_c$. Fig. 2 shows the $M(R)$ dependence for
different values of the bag parameter $B$. It is seen that the
purely nucleon EOS leads to a maximum mass of $\sim
2.2~M_{\odot}$. The advent of a quark phase in the NS diminishes
this value up to $\sim 1.75~M_{\odot}$. According to [16], in the
case $\lambda>\lambda_{cr} =3/2$, an infinitesimal core of the new
phase is unstable. Our analysis shows that, for
$B<69.3$~MeV/fm$^3$, the density jump parameter has the values
$\lambda>\lambda_{cr}$, and NS configurations with infinitesimal
quark cores will be unstable. In the latter case, there is a
nonzero minimum value of the quark core radius of a neutron star.
Accretion of matter on a critical neutron star will then result in
a catastrophic (jumplike) transformation of the star, forming a
star with a quark core of finite size. The catastrophic
transformation with finite quark core formation at the stellar
center will release an enormous amount of energy, compared with
energy release at a Supernova explosion. Our calculations have
allowed an evaluation of this energy. Thus, for $B=60$~MeV/fm$^3$,
a star of mass $M\approx 0.24~M_{\odot}$ and radius $R\approx
16.77$ km forms, in a jump-like manner, a star of radius $R\approx
13.95$ km and a quark core, having a core mass
$M_{core}\approx0.087~M_{\odot}$ and radius $R_{core}\approx4.38$
km. The catastrophic restructuring process releases an energy
$E_{conv}\approx4~10^{43}$ J. For $B>90$ MeV/fm$^3$, the quark
degrees of freedom make the neutron star unstable, and
configurations with a quark core are absent. In the case $69.3\le
B\le 90$ MeV/fm$^3$, configurations with arbitrarily small quark
cores are stable.

\section{Conclusion}
In the framework of the RMF theory, we have considered the EOS of
NS matter and obtained that inclusion of the isovector-scalar
$\delta$-meson field leads to nonnegligible changes in the
characteristics of quark first-order phase transitions. For
different values of the bag parameter $B$, we have constructed
neutron star models with quark cores. The results have shown that,
depending on value of this parameter realized in the Nature, the
following criteria may take place. In the case $B>90$ MeV/fm$^3$,
the quark phase leads to a total neutron star instability, and
stars with quark cores are absent. In the case $69.3\le B\le 90$
MeV/fm$^3$, stable NS with arbitrarily small quark cores may
exist. If $B<69.3$ MeV/fm$^3$, then configurations with small
quark cores are unstable. In the latter case, accretion of matter
will result in a catastrophic NS transformation after which a
quark core of finite size will be formed.

\section*{Acknowledgements}{This work has been supported by the Ministry of
Education and Science of the Republic of Armenia, Grant 2008-130.}


\begin{thebibliography}{99}

\bibitem{Wal74} J. D. Walecka, Ann. Phys., \textbf{83}, 491 (1974).
\bibitem{SW97} B. D. Serot, J. D. Walecka, Int.J.Mod.Phys. \textbf{E6}, 515 (1997).
\bibitem{Typel99} S. Typel, H. H. Wolter, Nucl. Phys. \textbf{A656}, 331 (1999).
\bibitem{Mill95} H. Miller, B. D. Serot, Phys. Rev. \textbf{C52}, 2072 (1995).
\bibitem{Ko_Li96} C. M. Ko, G. Q. Li, Journal of Phys., \textbf{G22},1673 (1996).
\bibitem{Kubis97} S. Kubis, M. Kutschera, Phys. Lett., \textbf{B399},191 (1997).
\bibitem{Liu02} B. Liu, V. Greco, V. Baran, M. Colonna, M. Di Toro, Phys.Rev.
\textbf{C65}, 045201 (2002).
\bibitem{Greco03} V. Greco, M. Colonna, M. Di Toro, F. Matera, Phys. Rev. \textbf{C67},
015203 (2003).
\bibitem{Alav08} G. B. Alaverdyan, Astrofiz., \textbf{52}, 147, (2009).
\bibitem{Hof01} F. Hofmann, C. M. Keil, H. Lenske, Phys.Rev. \textbf{C64}, 034314
(2001).
\bibitem{BBP71} G. Baym, H. Bethe, Ch. Pethick, Nucl.Phys.,\textbf{ A175}, 255, 1971.
\bibitem{Chod74} A. Chodos, R. L. Jaffe, K. Johnson, C. B. Thorn, V. F. Weisskopf,
Phys.Rev.\textbf{D9}, 3471 (1974).
\bibitem{FarJaf84} E. Farhi, R. L. Jaffe, Phys.Rev. \textbf{D30}, 2379 (1984).
\bibitem{Glend92} N. K. Glendenning, Phys. Rev. \textbf{D 46}, 1274 (1992).
\bibitem{TOV39} R. Tolman, Phys.Rev. 55, 364, 1939; J. Oppenheimer and G. Volkoff,
Phys.Rev. \textbf{55}, 374 (1939).
\bibitem{Seid71} Z. N. Seidov, Astron. Zh., \textbf{15}, 347 (1971).

\end{thebibliography}
\end{document}